\documentclass[%
 reprint,
 amsmath,amssymb,
 aps,
prc
]{revtex4-2}

\usepackage{graphicx}
\usepackage{dcolumn}
\usepackage{bm}
\usepackage{hyperref}

\usepackage{color}
\usepackage{ulem}
\usepackage{multirow}
\usepackage{amsmath}
\usepackage{longtable}
\usepackage{CJK}
\usepackage{blindtext}
\usepackage{array}
\usepackage{tabularx}
\newcolumntype{M}[1]{>{\centering\arraybackslash}m{#1}}

\begin{document}

\begin{CJK*}{UTF8}{gbsn}


\title{Odd-even mass differences of well and rigidly deformed nuclei in the rare earth region: 
A test of a newly proposed fit of average pairing matrix elements}

\author{T. V. Nhan Hao}
\email{tvnhao@hueuni.edu.vn}
\affiliation{Faculty of Physics, University of Education, Hue University, 34 Le Loi Street, Hue City, Vietnam}
\affiliation{Center for Theoretical and Computational Physics, University of Education, Hue University, 34 Le Loi Street, Hue City, Vietnam}

\author{N. N. Bao Nguyen}
\affiliation{Faculty of Physics, University of Education, Hue University, 34 Le Loi Street, Hue City, Vietnam}
\affiliation{Center for Theoretical and Computational Physics, University of Education, Hue University, 34 Le Loi Street, Hue City, Vietnam}

\author{D. Quang Tam}
\affiliation{Faculty of Physics, University of Education, Hue University, 34 Le Loi Street, Hue City, Vietnam}
\affiliation{Faculty of Basic Sciences, University of Medicine and Pharmacy, Hue University, 06 Ngo Quyen Street, Hue City, Vietnam}

\author{P. Quentin}%
 \email{quentin@cenbg.in2p3.fr}
\affiliation{LP2I, UMR 5797, Universit\'{e} de Bordeaux, CNRS, F-33170, Gradignan, France
}%

\author{Meng-Hock Koh (辜明福)}
 \email{kmhock@utm.my}
 \affiliation{Department of Physics, Faculty of Science, Universiti Teknologi Malaysia, 
        81310 Johor Bahru, Johor, Malaysia.}
\affiliation{UTM Centre for Industrial and Applied Mathematics, 81310 Johor Bahru, Johor, Malaysia}

\author{L. Bonneau}%
\affiliation{LP2I, UMR 5797, Universit\'{e} de Bordeaux, CNRS, F-33170, Gradignan, France
}%

\date{\today}

\begin{abstract}
We discuss a test of a recently proposed approach to determine average pairing matrix elements 
within a given interval of single-particle states (sp) around the Fermi level $\lambda$ 
as obtained in the so-called uniform gap method (UGM). 
It takes stock of the crucial role played  
by the averaged sp level density $\tilde{\rho}(e)$. 
These matrix elements are deduced within the UGM approach, 
from microscopically calculated $\tilde{\rho}(e)$ and gaps obtained from analytical formulae of a semi-classical nature. 
Two effects generally ignored in similar fits have been taken care of. 
They are: (a) the correction for a systematic bias in choosing to fit pairing gaps corresponding 
to equilibrium deformation solutions as discussed by M\"{o}ller and Nix [Nucl. Phys. A 476, 1 (1992)] and 
(b) the correction for a systematic spurious enhancement of $\tilde{\rho}(e)$ for protons in the vicinity of $\lambda$, 
because of the local Slater approximation used for the treatment of the Coulomb exchange terms in most calculations 
(see e.g. [Phys. Rev C 84, 014310 (2011)]). 
This approach has been deemed to be very efficient upon performing Hartree-Fock + BCS 
(with seniority force and self-consistent blocking when dealing with odd nuclei) calculations 
of a large sample of well and rigidly deformed even-even rare-earth nuclei. 
The reproduction of their experimental moments of inertia has been found to be at least 
of the same quality as what has been obtained in a direct fit of these data [Phys. Rev C 99, 064306 (2019)]. 
We extend here the test of our approach to the reproduction, in the same region, 
of three-point odd-even mass differences centered on odd-$N$ or odd-$Z$ nuclei. 
The agreement with the data is again roughly of the same quality as what has been obtained in a direct fit, 
as performed in [Phys. Rev C 99, 064306 (2019)].
\end{abstract}

\maketitle
\end{CJK*}


\section{\label{sec:intro} Introduction}
A simple prescription to determine average pairing matrix elements from averaged single-particle (sp) level densities 
for the ground states of well and rigidly deformed nuclei has been proposed recently in \cite{MHKPQ} and deemed to be rather successful. 
These matrix elements $V_q$ (where $q$ stands for the charge state i.e. neutron or proton) 
correspond to their average values around the Fermi energies $\lambda_q$. 
They are to be used in a microscopic approach of the Hartree-Fock-plus-BCS (HF + BCS) type. 
The pairing correlations are treated within the so-called seniority force approximation.
It  consists in solving the BCS variational equations in a restricted sp space around the Fermi energy, 
assuming the constancy of the pairing matrix element within this interval.

It takes stock of the strong dependence of the $V_q$ values upon the sp level densities at the Fermi energies $g_q(\lambda_q)$ 
averaged \`{a} la Strutinsky considered as a sound ersatz of a semi-classical approximation. 
It  relies also on the consideration that any fitting of nuclear energies (here any energy related to pairing correlations) 
in terms of nucleon numbers correspond in an effective way to such a semi-classical approximation. 
Therefore the input of experimental data is naturally performed through some standard analytical formula 
expressing the nucleon number dependence of odd-even mass differences defined 
in different ways noted here as $\delta E$ (see, e.g., Refs.~\cite{Jensen1984,Madland1988}) 
taking into account the important corrective contribution of \cite{Moller1992}.

After the pionneering work of Ref.~\cite{BMP}, 
it has been a current practice to adjust the intensity of pairing correlations to reproduce the data on $\delta E$. 
To a lesser extent, for this purpose, one has also considered moments of inertia noted as $\mathcal{J}$, 
deduced from the first $2^{+}$ excitation energies of well and rigidly deformed even-even heavy nuclei. 
In a recent paper \cite{Hafiza2019}, it has been demonstrated in the region of (or close to) rare-earth deformed nuclei 
(loosely dubbed as rare-earth nuclei in what follows) that making separate fits 
on $\delta E$ and $\mathcal{J}$ yielded similarly good results for the parameters 
in use to calculate pairing correlations within HF + (seniority) BCS calculations. 
This was a confirmation that both pieces of data were highly contingent upon a good description 
of pairing properties and thus asserting their relevance for such fits. 

In both cases (with $\delta E$ or $\mathcal{J}$), these fitting approaches possibly 
suffer from accidental local deficiencies of the sp distribution of levels in the vicinity of the Fermi energy. 
\textit{A priori}, resorting to semi-classical quantities associated to a given microscopic theoretical approach 
one gets rid of such a practical problem. 
What remains of course to be assessed is the relevance of the analytical expressions for the $(N, Z)$ dependence 
of the average $\delta E$ differences as well as the quality of the averaged sp level density of the canonical basis states.

The method proposed in \cite{MHKPQ} has been validated by comparing its results for moments of inertia of 
well and rigidly deformed rare-earth nuclei with those obtained
through a specific fit on the $\mathcal{J}$ data 
as performed in Ref.~\cite{Hafiza2019}. 
A similar quality in the reproduction of such spectroscopic data was
obtained.

It is the aim of the present paper to make a similar comparison for the $\delta E$ data now, 
between the results obtained within the method developed in \cite{MHKPQ} and 
the direct fit of these differences performed in the same nuclear region, in Ref.~\cite{Hafiza2019}.

In Section~\ref{sec:the approach}, we will briefly recall the approach developed in \cite{MHKPQ}, 
whereas Section~\ref{sec:calculations details} will provide some details about the calculations performed here. 
Our results will be discussed in Section~\ref{sec:results} ; Section~\ref{sec:conclusion} will be devoted to some conclusions and perspectives.

\section{\label{sec:the approach} Brief survey of the approach}

The method in use has been presented and discussed in detail in \cite{MHKPQ}. 
We will thus restrict here to sketchily describe its main characteristics.

Our starting point is a sp spectrum for a charge state $q$, obtained \textit{a priori} through any microscopic approach. 
In what follows it will result from self-consistent
HF+BCS calculations within the seniority force simple ansatz 
for the pairing matrix elements. 

We calculate an approximate semi-classically averaged sp level density $\tilde{\rho}(e)$ 
as a function of the sp energy $e$ through a standard Strutinsky's energy averaging (see Refs. \cite{FunnyHill,NPA207})
using the equation:
\begin{equation}
    \tilde{\rho}_q (e) = \frac{1}{\gamma} \int_{-\infty}^{\infty} \rho(e') f\Big( \frac{e' - e}{\gamma}\Big) de'.
    \label{eq:average level density}
\end{equation}
As discussed in \cite{MHKPQ} for the considered nuclei located far enough from the neutron drip line, 
the averaging width is taken as $ \gamma = 1.2 \: \hbar \omega$ where the energy scale as a function of the nucleon number $A$ 
is the usual expression $\hbar \omega = 41 A^{- 1/3}$ MeV \cite{Moszkowski}. 
The $f(x)$ term corresponding to the so called curvature correction, defined as
\begin{equation}
    f(x) = P(x) \: w(x)
\end{equation} 
is a product of the weight factor $w(x)$
\begin{equation}
    w(x) = \frac{1}{\sqrt{\pi}} e^{-x^2}.
\end{equation}
and a polynomial $P(x)$ taken here to be the generalized Laguerre polynomials $L_{M}^{(\alpha)}$ 
for the variable $x^2$ of order $M =2$ written as
\begin{equation}
    P(x) = L_M^{1/2}(x^2) = \sum_{n = 0}^M a_{2n} \: x^{2n}.
    \label{eq: polynomial eq}
\end{equation}
with coefficients $\alpha_{2n}$ given e.g. in Table II of~\cite{MHKPQ}.

For $N_q$ nucleons of the charge state $q$, we then compute the Fermi energy $\tilde{\lambda}_q$ as
\begin{equation}
N_q  = \int_{-\infty}^{\tilde{\lambda}_q} \tilde{\rho}_q(e) de.
\end{equation}

Within the so-called uniform gap method of Refs. \cite{FunnyHill} \cite{RingSchuck}, 
given a suitably average gap $\tilde{\Delta}_q$ (see below), 
the average pairing matrix element $V_q$ to be used in our HF+BCS approach is given by
\begin{equation}
\frac{1}{V_q}  = \int_{\tilde{\lambda}_q - \Omega}^{\tilde{\lambda}_q + \Omega} 
\frac{\tilde{\rho}_q(e)}{\sqrt{(e - \tilde{\lambda}_q)^2 +{\tilde{\Delta}}_q^2}} de.
\label{eq: integration of ave density for pairing matrix element}
\end{equation}
As well-known, the value of this matrix element depends on the choice
of the energy interval $ 2 \Omega$ centered at the Fermi energy and
including the sp states active in the BCS variational determination of occupation probabilities. 
In this study as in \cite{MHKPQ} we take $\Omega = 6$ MeV.

The nucleon-number $N_q$ dependence of average gaps $\tilde{\Delta}_q$ could be taken
\textit{a priori} from standard formulas 
(as, e.g., those of Refs. \cite{Jensen1984}, \cite{Madland1988}). 
However M\"{o}ller and Nix \cite{Moller1992} have pointed out that there is a bias in such estimates 
due to the mere selection of nuclei at equilibrium deformation, 
corresponding systematically to lower than average quantal sp level densities. 
As a result of their study, they proposed the following parametrisation of the average gaps
\begin{equation}
    \tilde{\Delta}_{q} = \frac{r B_s}{N_q^{1/3}}\,,
    \label{eq: Moller Nix pairing gap}
\end{equation}
where $B_s$ is set to $1$ and $r = 4.8$ MeV.

While we stick to this value for neutrons, we have remarked in \cite{MHKPQ} that this must not be the case for protons. 
In most of microscopic approaches of the Hartree-Fock plus BCS,
Hartree-Fock-Bogoliubov (HFB) or relativistic mean-field type, 
one simplifies the treatment of exchange terms of the infinite range Coulomb interaction 
by having recourse to the local Slater approximation \cite{Slater}. 
It has been found long ago \cite{Titin} and further studied in Refs. \cite{Skalski,Bloas} 
that the Slater approximation systematically and significantly 
overestimates the quantal sp level density near the Fermi energy
(especially at the equilibrium deformation).

To ensure a safe use of the Slater approximation in our approach, 
we must thus quench the parameter $r$ of Eq~(\ref{eq: Moller Nix pairing gap}) by a 
factor $R_p$ 
which is the ratio of the pairing gaps obtained in two separate quantal BCS calculations 
(exact and approximated \`{a} la Slater). 
It has been shown in Ref.~\cite{MHKPQ} that the dependence of this ratio 
with respect to the intensity of pairing correlations may be approximatively given by 
\begin{equation}
    R_p = 0.0181 \: E_{cond}^p + 0.781\,,
    \label{eq:Rp correction factor}
\end{equation}
where $E_{cond}^p$ is the average pairing 
condensation energy (defined as the part of the total energy involving explicitely the abnormal BCS density) given in MeV.

It is estimated  using the quantal gap obtained within HF+BCS calculations 
with some starting ansatz for the proton matrix element $V_p$ such that
(see Appendix A of \cite{MHKPQ}):
\begin{equation}
    E_{cond}^p = \frac{\Delta_p^2}{V_p}
    \label{eq: condensation energy}
\end{equation}
where $\Delta_p$ is the BCS proton pairing gap (in MeV).

In principle this should imply an iterative process to define new M\"{o}ller-Nix average gaps 
and accordingly a new $V_q$ value. 
However, it has been shown in \cite{MHKPQ} that
choosing some starting $V_p$ 
which is approximated by a constant matrix element 
with an initial pairing strength $G_q = 19$ MeV (see Sec.~\ref{subsec: canonical basis})
range of values (as it turns out easily delineated), 
these iterations do not bring significant modifications of the resulting matrix elements 
given the rough character of the above corrective process.   
As such, we limit ourselves herein to a non-iterative process of estimating the pairing matrix elements.

\section{\label{sec:calculations details}Some details of the calculations}
\subsection{\label{subsec: canonical basis}The canonical basis}
In our approach, the sp canonical basis states are obtained within self-consistent HF+BCS calculations. 
Let us recall that throughout this paper the BCS correlations are determined 
within the seniority force simple approximation. 
As an alternative presentation for the values of the  the average matrix elements $V_q$ we provide $G_q$ parameters 
introduced in Ref. \cite{Bonche_pairing} to the effect of removing approximately 
the dependence of these matrix elements from $N_q$ values
\begin{equation}
    V_q = \frac{G_q}{11 + N_q}.
    \label{eq:constant G}
\end{equation}

The particle-hole interaction in use is of the standard Skyrme type.
Calculations are performed with the SIII parametrisation \cite{SIII} 
which has been shown in many instances to provide a rather good description 
of the spectroscopic properties of well and rigidly deformed nuclei (see, e.g., for a recent account Ref. \cite{K_isomer}).
The averaged pairing matrix elements $V_q$ are given for each nucleus according to the method summarized in Section~\ref{sec:the approach}
for even-even nuclei. 
For odd-$Z$ (odd-$N$ respectively) nuclei the retained values of the matrix elements 
are interpolated between the values found for the neighbouring isotones (isotopes respectively).

The axial and intrinsic parity symmetries have been imposed to the solutions. 
The nuclei are thus specified by their projection $K$ of their angular momentum on the symmetry axis
and their parity $\pi$.
To solve the Schr\"{o}dinger equation determining the canonical basis states, 
we have expanded these states on eigenstates of an axially symmetrical harmonic oscillator Hamiltonian. 
It is defined by a basis size parameter $N_0  = 16$ 
defining a deformation-dependent selection of relevant basis states 
corresponding at sphericity to the consideration of 17 shells.
This basis is truncated according to (see Ref. \cite{NPA203}) 
\begin{equation}
    \hbar \omega_\bot (n_\bot + 1) + \hbar \omega_z (n_z + \frac{1}{2} ) \le \hbar \omega_0 (N_0 + 2).
\end{equation}
where $n_z$ and $n_{\bot}$ are the number of oscillator quanta in the symmetry-axis direction ($z$-axis) 
and in the perpendicular direction and $N_0+1$ corresponds to the number of spherical shells.

The inverse-length $b$ and
deformation $q$ parameters are defined, $m$ being the average nucleonic mass,
such that \cite{NPA203}
\begin{equation}
    b = \sqrt{\frac{m \omega_0}{\hbar}} \;\;\; ; \;\;\; q = \frac{\omega_\bot}{\omega_z}.
\end{equation}
These two parameters are related to the angular frequencies on the $(x,y)$ plane, $\omega_\bot$, and $z$-axis, $\omega_z$,
while $\omega_0^3 = \omega_\bot^2 \omega_z$ is the angular frequency at sphericity.
The two parameters $b$ and $q$ have been
optimized for even-even nuclei
to give the lowest energy for each equilibrium solution,
whereas for odd-$A$ nuclei they have been interpolated
  from those obtained for even-even nuclei. 
Integrals involving the local densities are performed using the Gauss-Hermite and Gauss-Laguerre 
approximate integration methods with 50 points along the symmetry axis
  and 16 points in the perpendicular direction, respectively.

\subsection{\label{subsec: odd-even mass} Odd-even mass differences}

The data on odd-even mass differences are extracted through a 3-point formula. 
As discussed in Refs. \cite{consist12,consist13} such differences $\delta_q^{(3)}$
centered on an odd-neutron or odd-proton nucleus are good markers on
the degree of pairing correlations.
They are, to a large extent, free from single-particle filling effects
and are given, e.g., for an isotopic series by
\begin{flalign}
    \delta^3_n (N) &= \frac{(-1)^N}{2} \Bigg[E(N+1,Z) - 2E(N,Z) + E(N-1,Z)\Bigg] \notag \\
    &= \frac{(-1)^N}{2} \Bigg[S_n(N,Z) - S_n(N+1,Z)\Bigg]
    \label{eq:odd-even mass difference}
\end{flalign}
where $N$ is odd and $S_n(N,Z)$ is the neutron separation energy 
of a nucleus composed of $N$ neutrons and $Z$ protons
whose total energies are denoted as $E(N,Z)$. 
Similar expressions are easily deduced from the above for odd-proton nuclei.

These energies are compared with those extracted directly from calculated binding energies within the HF+BCS approach. 
In the case of odd-$A$ nuclei, we have performed selfconsistent blocking calculations 
(i.e. placing a nucleon in the relevant sp orbit specified by the $K^{\pi}$ quantum numbers). 
The breaking of the time-reversal symmetry implies the presence of new (time-odd) local densities 
in the expression of the Hamiltonian density resulting in new terms in the corresponding Hartree-Fock potential. 
As explicitely detailed in Ref. \cite{Magnet}, when using the SIII interaction we have considered a restricted choice 
of time-odd potential fields, yet preserving the Galilean invariance 
(namely the vector spin field $\bold S(\bold r)$ and the vector current field $\bold A(\bold r)$, with usual notation). 
This choice is dubbed as the \textit{minimal} scheme in Ref. \cite{Magnet}.

In the spirit of the Bohr-Mottelson unified model, well suited for these deformed nuclei, 
we assimilate the nuclear angular momentum and parity quantum numbers
$I^{\pi}$ to those $K^{\pi}$ of the blocked blocked-nucleon sp state. 
This is of course not free for any perturbation of the low-energy nuclear spectra from possible Coriolis coupling 
which will be ignored here. 
Particular cases concern solutions where $K = 1/2$ with a decoupling parameter $a$ outside 
the range $ - 1 \le a \le 4$. 
They will be specifically discussed in Section~\ref{sec:results}
and shown to be unable to perturb the \textit{natural} rotational band ordering of states.
The decoupling parameter $a$ is defined, as well known \cite{BM1}, through the relation
\begin{equation}
a = -\langle i |\widehat J_+ | \widetilde i \rangle
\end{equation}
where $\widehat J_+$ is the usual angular momentum ladder operator (sum of orbital and spin angular momenta) in $\hbar$ unit 
and $|\widetilde i\rangle$ is the single-particle state canonically conjugate of the blocked state $|i\rangle$. 
These two states are such that $\widehat J_z|i\rangle = \Omega_i |i\rangle$ 
and $\widehat J_z|\widetilde i\rangle = -\Omega_i |\widetilde i\rangle$. 
In our HFBCS code with selfconsistent blocking, the time-reversal symmetry is broken 
in the one-body sector and the definition 
of the two sp states forming the equivalent of Cooper pairs in our BCS wavefunction is provided in
Appendix A of \cite{Koh_epja_2016}.

We have systematically calculated solutions for odd-$A$ nuclei corresponding to the ground state $I^{\pi}$
experimental values \cite{NNDC} as well as those where the calculated energies lie below 
the former or in some cases (see the discussions of Section~\ref{sec:results}) above generally up to a couple of 100 keV.

\subsection{\label{subsec:choice of nuclei}Choice of sample nuclei}
We have chosen odd-$A$ nuclei belonging to the rare-earth region plus Hafnium isotopes (loosely called below rare-earth nuclei), 
i.e. some odd-proton isotopes from Europium to Lutetium and odd-neutron isotopes from Samarium to Hafnium. 
They are chosen to be well and rigidly deformed and, moreover, far enough from a region of transition between deformed and soft nuclei. 
The first criterion is retained to be able to reduce approximately the collective dynamics to a pure rigid rotation 
within the Bohr-Mottelson unified model, allowing in particular to limit ourselves to a single BCS state, i.e. 
ignoring quantal shape fluctuations. 
The second criterion is retained to avoid wide shape variations (and therefore large sp spectrum changes) 
between the three isotopes (isotones respectively) 
entering the calculation of the energy differences $\delta_n^{(3)}$ ($\delta_p^{(3)}$ respectively).

Table~\ref{tab:results even} displays the ratio $R_{42}$ of the
excitation energies of the first $4^{+}$ and $2^{+}$ states \cite{NNDC}
of the 22 even-even nuclei bracketing the odd-$A$ nuclei whose energy differences $\delta_q^{(3)}$ are evaluated in our calculations. 
It appears that they satisfy reasonably well the first criterion since for all of them one has $R_{42} \ge 3.29$. 

On this table also, the intrinsic axial charge quadrupole moments $Q^{\rm int}_{20}$ obtained in our calculations are compared, 
when available, with the corresponding experimental values deduced either from reduced B(E2) data \cite{BE2} 
or those deduced from the spectroscopic moments $Q^{\rm spect}_{20}$ of the first $2^{+}$ state \cite{Spect}, 
upon using the unified model relation for $I = 2, K = 0$, namely $Q^{\rm int}_{20} = - 3.5$ $Q^{\rm spect}_{20}$. 
One notices a rather good reproduction of the $Q^{\rm int}_{20}$ data with the interaction SIII in use, 
as shown long ago \cite{PLB} (somewhat less good however for the $^{156, 158}$Sm isotopes). 

	\begin{table*}[]
		\caption{Some static properties of the 22 even-even nuclei considered in this paper. 
            The calculated total binding energies $E^{\rm theory}$ are given in MeV. 
            The experimental intrinsic axial quadrupole moments for the charge distribution 
                are deduced (within the unified model relations for rotational band states) from 
            Refs.~\cite{Spect} (\cite{BE2} respectively) for spectroscopic first $2^{+}$ states data 
            noted as $Q_{20}^{\rm spect}$ moments (for moments deduced from reduced $B(E2)$ data noted as $Q_{20}^{\rm BE2}$ respectively). 
            They are compared with the corresponding calculated moments $Q_{20}^{\rm theory}$. All these moments are given in barn.
                }
            \label{tab:results even}
        \begin{ruledtabular}
        \renewcommand{\arraystretch}{1.5}
		\begin{tabular}{*{6}c}   
			Nucleus	& E$^{\text{theory}} $	&R$ _{42} $	&Q$ ^{\text{spect}}_{20} $	&Q$ ^{\text{BE2}}_{20} $	&Q$ ^{\text{theory}}_{20} $\\
			\hline
			$ ^{156} $Sm	&-1276.496&	3.290	&5.85 (7)	&   $-$&	6.81	\\
			$ ^{158} $Sm	&-1288.582&	3.301	&6.55 (14)	&  $-$ &   6.99			\\
			$ ^{160} $Sm	&-1299.953&	3.292	&$-$  &   $-$&		7.11\\
			$ ^{160} $Gd	&-1305.533&	3.302	&7.28 (14)&	7.265 (42)&	7.25\\
			$ ^{162} $Gd	&-1318.251&	3.302	& $-$&    $-$&		7.40\\
			$ ^{164} $Gd	&-1330.079&	3 295	& $-$&    $-$&	7.51	\\
			$ ^{166} $Gd	&-1341.067&	3.300	& $-$&    $-$&	7.58	\\
			$ ^{162} $Dy	&-1319.955&	3.294	& $-$&	7.33  (8)	&7.38\\
			$ ^{164} $Dy	&-1333.969&	3.301	&7.28  (53)	&7.503 (33)	&7.56\\
			$ ^{166} $Dy	&-1347.128&	3.310	& $-$&    $-$&		7.69\\
			$ ^{168} $Dy	&-1359.411&	3.313	& $-$&    $-$&		7.76\\
			$ ^{168} $Er	&-1361.733&3.309	&  $-$&    7.63 (7)	&7.84\\
			$ ^{170} $Er	&-1375.231&	3.310	&6.65 (70)	&7.65 (7)	&7.93\\
			$ ^{172} $Er	&-1387.702&	3.314	& $-$&    $-$&		7.72\\
			$ ^{170} $Yb	&-1374.077&	3.293	&7.63 (11)	&7.63 (9)  	&7.90\\
			$ ^{172} $Yb	&-1388.724&	3.305	&7.77 (14)	&7.792 (45)	&7.98\\
			$ ^{174} $Yb	&-1402.460&	3.310	&7.63 (18)	&7.727 (39)	&7.77\\
			$ ^{176} $Yb	&-1415.595&	3.310	&7.98 (21)	&7.30 (13)	&7.58\\
			$ ^{178} $Yb	&-1427.957&	3.310	& $-$&    $-$&		7.46\\
			$ ^{178} $Hf	&-1429.290&	3.291	&7.07  (7)	&6.961 (43)	&7.22\\
			$ ^{180} $Hf	&-1442.851&	3.307	&7.00  (7)	&6.85  (9)	&7.08\\
			$ ^{182} $Hf	&-1455.193&	3.295	& $-$&    $-$&		6.85\\
		\end{tabular}
        \end{ruledtabular}
	\end{table*}

\section{\label{sec:results} Results}

While this paper is concerned with a test of calculated odd-even mass differences $\delta_q^{(3)}$,  
we will first discuss the nature and relevance of the configurations retained in our comparison with experimental data. 
By configuration we mean the nuclear spin 
(assumed as we have seen, to be equal to the projection of the angular momentum on the quantification axis) 
and parity (well defined in our solution since we impose an intrinsic reflection symmetry). 
Obviously the choice of the configuration determining the location in the sp spectrum of the unpaired nucleon 
has a direct effect on the relevant separation energies. 
As a rule, however, to avoid overestimating unduly the predictive value of our approach, 
we will retain in our comparison with the data, theoretical solutions possessing the experimental ground state spin and parity values 
even though they do not correspond to the lowest calculated total energy. 
It was nevertheless interesting to see how well the spin and parity of the calculated ground states match with the data. 
In the same way, since the ordering of sp states around the Fermi depends significantly on the deformation of the mean field, 
we checked the agreement of our calculated axial moments with the intrinsic moments extracted from two pieces 
of experimental data (reduced E2 transition and spectroscopic moment data).

\subsection{Discussion of the ground-state configurations obtained in our calculations for odd-Z nuclei}

We compare here the lowest-energy configurations as obtained in our calculations with the 
experimental values of the angular momentum and parity quantum numbers $I^{\pi}$ given in the 
current version of the NUDAT compilation \cite{NNDC}.

As seen on Table~\ref{tab:result odd-proton}, for 8 nuclei out of 13 
($^{161}$Tb, $^{163}$Tb, $^{167}$Ho, $^{169}$Ho, $^{169}$Tm, $^{171}$Tm, $^{177}$Lu, $^{179}$Lu), 
our theoretical assignments agree with the data.

We confirm the suggested assignments for 3 nuclei, namely: $^{165}$Tb and $^{167}$Tb as $3/2^{+}$ and $^{173}$Tm as $1/2^{+}$.
No assignment is proposed in Ref. \cite{NNDC} for the $^{161}$Eu nucleus. 
We suggest a $5/2^{-}$ configuration noting however that we have
obtained a $5/2^{+}$ solution $179$ keV above the $5/2^-$ configuration.
In one case ($^{159}$Eu) the experimental lowest configuration
$5/2^{+}$ is obtained at 145 keV above a  $5/2^{-}$ state.
Finally, we remark that the decoupling constant values corresponding to the $1/2^{+}$ state considered 
in the three calculated isotopes of Thulium belong to the $[-0.64, -0.60]$ interval ensuring 
that the band head spin assignment as $1/2$ is correct. 

    \begin{table*}
		\caption{Comparison of some spectroscopic properties of the 13 odd-$Z$ nuclei considered in this paper. 
            Along with the calculated total binding energies $E^{\rm theory}$ given in MeV, 
            the intrinsic configuration spins and parities $K^{\pi}$ of our solutions are reported. 
            The corresponding experimental ground state values $I^{\pi}$  (proposed or suggested - in brackets -  when available) 
            for a given nucleus are displayed for the sake of comparison, 
            assuming the validity of the unified model assumption of $I = K$ for the band head states. 
            Both calculated $\delta_{\rm calc}^{(3)}$ and experimental $\delta_{\rm exp}^{(3)}$ 
            odd-even mass differences (given in keV) are also reported.
                }
        \label{tab:result odd-proton}
        \begin{ruledtabular}
        \renewcommand{\arraystretch}{1.5}
		\begin{tabular}{*{6}c}   
			Nucleus	&$ \text{I}^{\pi}(\text{exp}) $	&$ \text{I}^{\pi}(\text{calc}) $	
                &$ \text{E}^{\text{theory}} $	&$ \delta^{(3)}_{\text{calc}} $	&$ \delta^{(3)}_{\text{exp}} $\\
			\hline
			$ ^{159} $Eu	&5/2$ ^{+} $	&		5/2$ ^{+} $&-1296.280&777 &554\\
							&					&	5/2$ ^{-} $ &-1296.425&633	& $-$\\
            \colrule
			$ ^{161} $Eu	& 					$-$&	5/2$ ^{-} $ &-1308.524&579 &  $-$\\
						&						&5/2$ ^{+} $	&-1308.345&757	&466\\
            \colrule
			$ ^{161} $Tb&	3/2$ ^{+} $	&3/2$ ^{+} $&-1312.161&583&	600\\
            \colrule
			$ ^{163} $Tb&	3/2$ ^{+} $	&3/2$ ^{+} $&-1325.597&513&	529\\
            \colrule
			$ ^{165} $Tb&	(3/2$ ^{+} $)&	3/2$ ^{+} $&-1338.149&455&	550\\
			\colrule
            $ ^{167} $Tb&	(3/2$ ^{+} $)&	3/2$ ^{+} $&-1349.801&438&	579\\
            \colrule
			$ ^{167} $Ho&	7/2$ ^{-} $&	7/2$ ^{-} $&-1353.783&648&	508\\
            \colrule
			$ ^{169} $Ho&	7/2$ ^{-} $&	7/2$ ^{-} $&-1366.734&588&	535\\
            \colrule
			$ ^{169} $Tm&	1/2$ ^{+} $&	1/2$ ^{+} $&-1367.174&732&	602\\
            \colrule
			$ ^{171} $Tm&	1/2$ ^{+} $&	1/2$ ^{+} $&-1381.392&585&	471\\
            \colrule
			$ ^{173} $Tm&	(1/2$ ^{+} $)&	1/2$ ^{+} $&-1394.585&496&	458\\
            \colrule
			$ ^{177} $Lu&	7/2$ ^{+} $&	7/2$ ^{+} $&-1422.045&397&	579\\
            \colrule
			$ ^{179} $Lu&	7/2$ ^{+} $&	7/2$ ^{+} $&-1435.055&348&	669\\
		\end{tabular}
        \end{ruledtabular}
	\end{table*}

	\begin{table*}[]
		\caption{Same as Table~\ref{tab:result odd-proton} for the 16 odd-$N$ nuclei considered in this paper.
                Note that for the $^{165}$Dy and $^{171}$Yb nuclei, 
                the experimental values of the spin and parity of a low-lying isomeric state have been also displayed.
                }
            \label{tab:result odd-neutron}
        \begin{ruledtabular}
        \renewcommand{\arraystretch}{1.5}
		\begin{tabular}{*{6}c}   
			Nucleus	&$ \text{I}^{\pi}(\text{exp}) $	&$ \text{I}^{\pi}(\text{calc}) $&   
                $ \text{E}^{\text{theory}} $	&$ \delta^{(3)}_{\text{calc}} $	&$ \delta^{(3)}_{\text{exp}} $\\
			\hline
			$ ^{157} $Sm&$ (3/2^{-}) $	&3/2$ ^{-} $&	-1281.764&	775&	629\\
            \colrule
			$ ^{159} $Sm&5/2$ ^{-} $	&5/2$ ^{-} $&	-1293.550&	718&	535\\
            \colrule
			$ ^{161} $Gd	&5/2$ ^{-} $&5/2$ ^{-} $&	-1311.129&	763&	605\\
            \colrule
			\multirow{3}{*}{$^{163}$ Gd}& \multirow{3}{*}{($5/2^{-},7/2^{+}$)}&    7/2$ ^{+} $&	-1323.398&	766&	\multirow{3}{*}{599}\\
			&&5/2$ ^{-} $&	-1322.819&	1346&	 \\
			&&1/2$ ^{-} $&	-1323.457&	707&	\\
            \colrule
			$ ^{165} $Gd&    $-$&    7/2$ ^{+} $&	-1334.916&	657&507	\\
            \colrule
			$ ^{163} $Dy&	5/2$ ^{-} $	&5/2$ ^{-} $&	-1326.097&	864&	694\\
            \colrule
			$ ^{165} $Dy&	$ 7/2^{+},1/2^{-}_m $&7/2$ ^{+} $&	-1339.788&	760	&664\\
			&&1/2$ ^{-} $&	-1339.836&	712&	$-$\\
            \colrule
			$ ^{167} $Dy&	$ (1/2^{-})$&1/2$ ^{-} $&	-1352.563&	707&	661\\
			&&7/2$ ^{+} $&	-1352.620&	650	&  $-$\\
            \colrule
			$ ^{169} $Er&	1/2$ ^{-} $	&1/2$ ^{-} $&	-1367.745&	737&	627\\
			&&7/2$ ^{+} $&	-1367.817&	665&	$-$\\
            \colrule
			$ ^{171} $Er&	5/2$ ^{-} $	&5/2$ ^{-} $&	-1380.826&	640&	577\\
            \colrule
			$ ^{171} $Yb	&$ 1/2^{-},7/2^{+}_m $&1/2$ ^{-} $&	-1380.625&	776&	703\\
			&&7/2+&	-1380.698&	703&	$-$\\                     
            \colrule
			$ ^{173} $Yb	&5/2$ ^{-} $&5/2$ ^{-} $&	-1394.879&	713&	549\\
            \colrule
			$ ^{175} $Yb&	7/2$ ^{-} $&7/2$ ^{-} $&	-1408.299&	728&	522\\
            \colrule
			$ ^{177} $Yb&	9/2$ ^{+} $&9/2$ ^{+} $&	-1421.235&	541&	598\\
            \colrule
			$ ^{179} $Hf&	9/2$ ^{+} $&9/2$ ^{+} $&	-1435.547&	524&	644\\
            \colrule
			$ ^{181} $Hf&  1/2$ ^- $& 1/2$ ^- $ &  -1448.366&  655&  512\\
		\end{tabular}
        \end{ruledtabular}
	\end{table*}
 
\subsection{Discussion of the ground-state configurations obtained in our calculations for odd-$N$ nuclei}

As seen on Table~\ref{tab:result odd-neutron}, for 9 nuclei out of 16 
($^{159}$Sm, $^{161}$Gd, $^{163}$Dy, $^{171}$Er, $^{173}$Yb, $^{175}$Yb, $^{177}$Yb, $^{179}$Hf, $^{181}$Hf), 
our theoretical assignments agree with the data.

We confirm the suggested assignment for one nucleus, namely: $^{157}$Sm as $3/2^{-}$.
No assignment is proposed in Ref.~\cite{NNDC} for the $^{165}$Gd nucleus. 
We suggest a $7/2^{+}$ configuration.
In $^{167}$Dy the suggested ground state configuration $1/2^{-}$ suggested in Ref.~\cite{NNDC} is
calculated only 58 keV above a $7/2^{+}$ state. 
In one nucleus ($^{169}$Er) the experimental lowest configuration
$1/2^{-}$ is obtained only 72 keV above a $7/2^{+}$ state.  

Three cases deserve a particular attention. 
In Ref.~\cite{NNDC} two assignments ($5/2^{-}, 7/2^{+}$) are suggested for $^{163}$Gd. 
We found the latter ($7/2^{+}$) as the ground state with
the former ($5/2^{-}$) lying 579 keV higher with a $1/2^{-}$ configuration located in between at 59 keV. 
In $^{165}$Dy one has found experimentally a $7/2^{+}$ ground state with a $1/2^{-}$ isomeric state 
with an excitation energy of 108 keV.
We found both states as the lowest ones but with an inversion, the $1/2^{-}$ being found 48 keV lower in energy.
A similar situation is also present in $^{171}$Yb with an experimental $1/2^{-}$ ground state 
and a $7/2^{+}$ isomeric state 95 keV above. 
Here, too, we found these states as the lowest ones but with an inversion whereby the $1/2^{\-}$ being found 73 keV lower.

\subsection{Comments on our assignments of spin and parity}
In our calculations for both odd-$Z$ and odd-$N$ nuclei, we thus found that we are in agreement in 21 out 29 cases 
where an assignment of spin and parity has been reported or suggested in Ref. \cite{NNDC}. 
We have proposed an assignment for two nuclei. 
In two instances where an isomeric state has been experimentally found at excitation energies in the 100-150 keV range,
we have obtained them, yet with an inversion in their respective ordering, corresponding to an error of 50-70 keV. 
From all these results implying a sample of 29 odd-$A$ nuclei, 
yet limiting ourselves to the consideration of intrinsic states within the unified model, 
we can reasonably hint that our estimates of the low lying bandhead spectra 
provides a good reproduction of relative energies within not much more than about 150 keV.

Finally we discuss the decoupling constant values corresponding to the $1/2$ states considered in our calculations of these odd-$N$ nuclei. 
It is found to be equal to $- 0.61$ for the $1/2^{-}$ state of $^{163}$Gd, $- 0.64$ and  $- 0.65$ for the $1/2^{-}$ states of $^{165}$Dy 
and $^{167}$Dy, $- 0.73$ for the $1/2^{-}$ state of 
$^{169}$Er, $- 0.76 $ for the $1/2^{-}$ state of $^{171}$Yb and  $+ 0.24 $ for the $1/2^{-}$ state of $^{181}$Hf.
In all cases, such values warrants that the bandhead spin assignment as $1/2$ is correct.

\subsection{Discussion of the odd-even mass differences obtained in our calculations}

Tables~\ref{tab:result odd-proton} and \ref{tab:result odd-neutron} display 
the three-points odd-even mass differences $\delta_q^{(3)}$ for protons and neutrons. 
The rms differences between experimental (as deduced from the separation energies given in Ref. \cite{NNDC}) 
and calculated in all cases for the experimental $I^{\pi}$ assignments are 165 keV for protons and 141 keV for neutrons. 
We can compare these values with what has been obtained in Ref.~\cite{Hafiza2019} 
where direct fits of the pairing matrix elements $V_n$ and $V_p$ have been performed 
on the odd-even mass differences $\delta_n^{(3)}$ and $\delta_p^{(3)}$ for a similar sampling of nuclei. 
There, the rms deviations were calculated as 182 keV for protons and 78 keV for neutrons. 
Given the expected accuracy on energies of our approach, 
we can deem the quality of our current approach to be  comparable to those of Ref. \cite{NNDC}.

To conclude this Subsection, it is instructive to assess numerically on two test cases, 
the effect of the two improvements with respect to the standard fitting procedure which have been considered in this paper 
(the M\"{o}ller-Nix prescription and the correction for the deficiency of the Slater approximation)

We display in Table~\ref{tab:slater correction test} the average matrix elements $V_q$ and the resulting odd-even mass-differences
with and without taking the $R_p$ corrective factor for two odd-mass nuclei
and consider two types
of empirical formula, namely that of Jensen and that of M\"{o}ller-Nix.

We first compare the results obtained
using the Jensen and M\"{o}ller-Nix formulas without Slater correction.
Using the Jensen pairing gaps, the absolute values
of the estimated neutron and proton pairing matrix elements
are always lower than in the M\"{o}ller-Nix case. This decrease of about 10 keV
in both neutron and proton pairing matrix elements
is translated into a decrease of the odd-even mass differences of about 200 keV.

For the odd-neutron $^{177}$Yb nucleus and the neighbouring even-even nuclei, 
using the M\"{o}ller-Nix gaps,
inclusion of the $R_p$ corrective factor 
decreases the absolute value of the proton pairing matrix elements (i.e. less pairing) by about 10 keV,
having no significant effect, as expected, on the neutron odd-even mass difference around the $^{177}$Yb nucleus.

For the odd-proton $^{177}$Lu nucleus, including the Slater correction 
yields also a decrease of about 10 keV of the average pairing matrix element resulting 
in a decrease of the calculated $\delta_p^3$ of about 250 keV.

\begin{table*}[]
    \caption{Absolute values of the average neutron ($V_n$) and proton ($V_p$) pairing matrix elements 
        together with the calculated binding energies (BE)
        and odd-even mass differences $\delta^3_q$ obtained with four different approaches in
        determining the pairing gap $\Delta_q$ for the $^{177}$Yb (odd-neutron with $K^\pi = 9/2^+$) 
        and $^{177}$Lu (odd-proton with $K^\pi = 7/2^+$) nuclei.
        Two types of empirical formulas are considered namely the Jensen and M\"{o}ller-Nix
        in which the formula are either used as they are, or in which the Slater correction is taken into account
        via equation~(\ref{eq:Rp correction factor}).}
        \label{tab:slater correction test}
    \begin{ruledtabular}
    \begin{tabular}{*{9}c}
    Nucleus&    Gap formula&  $Z$&    $N$&    $A$&    $V_n$ [MeV]&  $V_p$ [MeV]&  BE [MeV]&   $\delta^3_q$ [keV]  \\
    \hline
    \multirow{12}{*}{$^{177}$Yb}&  \multirow{3}{*}{Jensen}&    
        70	&	106	&	176	&	0.1582	&	0.2221	&	-1415.378	&		\\
    &&   70	&	107	&	177	&	0.1553	&	0.2205	&	-1421.230	&	351	\\
    &&   70	&	108	&	178	&	0.1524	&	0.2188	&	-1427.784	&		\\
	\cline{2-9}
    &   \multirow{3}{*}{Jensen + Slater corr.}&   
        70	&	106	&	176	&	0.1582	&	0.2093	&	-1415.237	&		\\
    &&   70	&	107	&	177	&	0.1553	&	0.2075	&	-1421.110	&	401	\\
    &&   70	&	108	&	178	&	0.1524	&	0.2057	&	-1427.784	&		\\
	\cline{2-9}									
    &   \multirow{3}{*}{M\"{o}ller-Nix}&    
        70	&	106	&	176	&	0.1681	&	0.2373	&	-1415.848	&		\\
    &&   70	&	107	&	177	&	0.1664	&	0.2364	&	-1421.462	&	544	\\
    &&   70	&	108	&	178	&	0.1646	&	0.2355	&	-1428.164	&		\\
	\cline{2-9}									
    &   \multirow{3}{*}{M\"{o}ller-Nix + Slater corr.}&    
        70	&	106	&	176	&	0.1681	&	0.2215	&	-1415.595	&		\\
    &&   70	&	107	&	177	&	0.1664	&	0.2210	&	-1421.235	&	541	\\
    &&   70	&	108	&	178	&	0.1646	&	0.2204	&	-1427.957	&		\\
    \hline  
    \multirow{12}{*}{$^{177}$Lu} &	\multirow{3}{*}{Jensen}& 
        70	&	106	&	176	&	0.1582	&	0.2221	&	-1415.378	&	\\
    &&	71	&	106	&	177	&	0.1603	&	0.2210	&	-1421.858	&	432     \\
    &&	72	&	106	&	178	&	0.1623	&	0.2199	&	-1429.201	&	\\
	\cline{2-9}												
    &	\multirow{3}{*}{Jensen + Slater corr.}&  
        70	&	106	&	176	&	0.1582	&	0.2093	&	-1415.237	&	\\
    &&	71	&	106	&	177	&	0.1603	&	0.2088	&	-1421.858	&	270    \\
    &&	72	&	106	&	178	&	0.1623	&	0.2082	&	-1429.017	&	\\
	\cline{2-9}										
    &  \multirow{3}{*}{M\"{o}ller-Nix}& 	
        70	&	106	&	176	&	0.1681	&	0.2373	&	-1415.848	&	\\
    &&	71	&	106	&	177	&	0.1683	&	0.2333	&	-1422.065	&	648 \\
    &&	72	&	106	&	178	&	0.1685	&	0.2292	&	-1429.579	&	\\
	\cline{2-9}										
    &   \multirow{3}{*}{M\"{o}ller-Nix + Slater corr.}&    
        70	&	106	&	176	&	0.1681	&	0.2215	&	-1415.595	&	\\
    &&	71	&	106	&	177	&	0.1683	&	0.2190	&	-1422.045	&	397    \\
    &&	72	&	106	&	178	&	0.1685	&	0.2164	&	-1429.290	&	\\
    \end{tabular}
\end{ruledtabular}
\end{table*}

\section{\label{sec:conclusion} Concluding remarks}
The present study complements in the rare-earth region and for the SIII Skyrme interaction what had been found in Ref.~\cite{MHKPQ}. 
There it was shown that the rms deviation between calculated and experimental moments of inertia deduced 
from the first $2^{+}$ level energy found in 11 well and rigidly deformed rare-earth nuclei was equal to 1.77 $\hbar^2 \mbox{MeV}^{-1}$ 
corresponding to about $5 \%$. 
It was almost equal (1.75  $\hbar^2 \mbox{MeV}^{-1}$) with what had been found from a direct fit of these moments of inertia 
(in the same region with the same interaction) in Ref.~\cite{Hafiza2019}.

Combining these results with ours,
we can conclude that the method proposed in \cite{MHKPQ} allows for a treatment of pairing correlations of the same quality 
as and simpler than a lengthy and more or less localized fit process to be performed for each particle-hole interaction.
However, it is \textit{a priori} suited to describe pairing correlations
only at the equilibrium deformation of well and rigidly deformed nuclei. 
In practice, all the calculations of the HF+BCS type within the seniority forces approach 
have assumed that the matrix elements obtained in these particular situations could be widely used 
in, e.g., computing potential-energy curves or surfaces as functions of some deformation parameters or multipole moments. 
Using the current approach as a starting point to determine the intensities of pairing residual interactions 
would free corresponding HF+BCS or HFB calculations from such ambiguities. This improvement is currently under study.

\begin{acknowledgments}
T. V. Nhan Hao, P. Quentin and D. Quang Tam acknowledge the support by the Hue University under the Core
Research Program, Grant no. NCM.DHH.2018.09.
Another co-author (M.H. Koh) would also like to acknowlege Universiti Teknologi Malaysia for its
UTMShine grant (grant number Q.J130000.2454.09G96).
\end{acknowledgments}

\bibliography{apssamp}

\begin{thebibliography}{27}%
\makeatletter
\providecommand \@ifxundefined [1]{%
 \@ifx{#1\undefined}
}%
\providecommand \@ifnum [1]{%
 \ifnum #1\expandafter \@firstoftwo
 \else \expandafter \@secondoftwo
 \fi
}%
\providecommand \@ifx [1]{%
 \ifx #1\expandafter \@firstoftwo
 \else \expandafter \@secondoftwo
 \fi
}%
\providecommand \natexlab [1]{#1}%
\providecommand \enquote  [1]{``#1''}%
\providecommand \bibnamefont  [1]{#1}%
\providecommand \bibfnamefont [1]{#1}%
\providecommand \citenamefont [1]{#1}%
\providecommand \href@noop [0]{\@secondoftwo}%
\providecommand \href [0]{\begingroup \@sanitize@url \@href}%
\providecommand \@href[1]{\@@startlink{#1}\@@href}%
\providecommand \@@href[1]{\endgroup#1\@@endlink}%
\providecommand \@sanitize@url [0]{\catcode `\\12\catcode `\$12\catcode `\&12\catcode `\#12\catcode `\^12\catcode `\_12\catcode `\%12\relax}%
\providecommand \@@startlink[1]{}%
\providecommand \@@endlink[0]{}%
\providecommand \url  [0]{\begingroup\@sanitize@url \@url }%
\providecommand \@url [1]{\endgroup\@href {#1}{\urlprefix }}%
\providecommand \urlprefix  [0]{URL }%
\providecommand \Eprint [0]{\href }%
\providecommand \doibase [0]{https://doi.org/}%
\providecommand \selectlanguage [0]{\@gobble}%
\providecommand \bibinfo  [0]{\@secondoftwo}%
\providecommand \bibfield  [0]{\@secondoftwo}%
\providecommand \translation [1]{[#1]}%
\providecommand \BibitemOpen [0]{}%
\providecommand \bibitemStop [0]{}%
\providecommand \bibitemNoStop [0]{.\EOS\space}%
\providecommand \EOS [0]{\spacefactor3000\relax}%
\providecommand \BibitemShut  [1]{\csname bibitem#1\endcsname}%
\let\auto@bib@innerbib\@empty
\bibitem [{\citenamefont {Koh}\ and\ \citenamefont {Quentin}(2024)}]{MHKPQ}%
  \BibitemOpen
  \bibfield  {author} {\bibinfo {author} {\bibfnamefont {M.-H.}\ \bibnamefont {Koh}}\ and\ \bibinfo {author} {\bibfnamefont {P.}~\bibnamefont {Quentin}},\ }\href@noop {} {\bibfield  {journal} {\bibinfo  {journal} {Phys. Rev. C}\ }\textbf {\bibinfo {volume} {110}},\ \bibinfo {pages} {024311} (\bibinfo {year} {2024})}\BibitemShut {NoStop}%
\bibitem [{\citenamefont {Jensen}\ and\ \citenamefont {Hansen}(1984)}]{Jensen1984}%
  \BibitemOpen
  \bibfield  {author} {\bibinfo {author} {\bibfnamefont {A.~S.}\ \bibnamefont {Jensen}}\ and\ \bibinfo {author} {\bibfnamefont {P.~G.}\ \bibnamefont {Hansen}},\ }\href@noop {} {\bibfield  {journal} {\bibinfo  {journal} {Nucl. Phys. A}\ }\textbf {\bibinfo {volume} {431}},\ \bibinfo {pages} {393} (\bibinfo {year} {1984})}\BibitemShut {NoStop}%
\bibitem [{\citenamefont {Madland}\ and\ \citenamefont {Nix}(1988)}]{Madland1988}%
  \BibitemOpen
  \bibfield  {author} {\bibinfo {author} {\bibfnamefont {D.}~\bibnamefont {Madland}}\ and\ \bibinfo {author} {\bibfnamefont {J.}~\bibnamefont {Nix}},\ }\href@noop {} {\bibfield  {journal} {\bibinfo  {journal} {Nucl. Phys. A}\ }\textbf {\bibinfo {volume} {476}},\ \bibinfo {pages} {1 } (\bibinfo {year} {1988})}\BibitemShut {NoStop}%
\bibitem [{\citenamefont {M{\"{o}}ller}\ and\ \citenamefont {Nix}(1992)}]{Moller1992}%
  \BibitemOpen
  \bibfield  {author} {\bibinfo {author} {\bibfnamefont {P.}~\bibnamefont {M{\"{o}}ller}}\ and\ \bibinfo {author} {\bibfnamefont {J.}~\bibnamefont {Nix}},\ }\href@noop {} {\bibfield  {journal} {\bibinfo  {journal} {Nucl. Phys. A}\ }\textbf {\bibinfo {volume} {536}},\ \bibinfo {pages} {20} (\bibinfo {year} {1992})}\BibitemShut {NoStop}%
\bibitem [{\citenamefont {Bohr}\ \emph {et~al.}(1958)\citenamefont {Bohr}, \citenamefont {Mottelson},\ and\ \citenamefont {Pines}}]{BMP}%
  \BibitemOpen
  \bibfield  {author} {\bibinfo {author} {\bibfnamefont {A.}~\bibnamefont {Bohr}}, \bibinfo {author} {\bibfnamefont {B.~R.}\ \bibnamefont {Mottelson}},\ and\ \bibinfo {author} {\bibfnamefont {D.}~\bibnamefont {Pines}},\ }\href@noop {} {\bibfield  {journal} {\bibinfo  {journal} {Phys. Rev.}\ }\textbf {\bibinfo {volume} {110}},\ \bibinfo {pages} {936} (\bibinfo {year} {1958})}\BibitemShut {NoStop}%
\bibitem [{\citenamefont {Nor}\ \emph {et~al.}(2019)\citenamefont {Nor}, \citenamefont {Rezle}, \citenamefont {Kelvin-Lee}, \citenamefont {Koh}, \citenamefont {Bonneau},\ and\ \citenamefont {Quentin}}]{Hafiza2019}%
  \BibitemOpen
  \bibfield  {author} {\bibinfo {author} {\bibfnamefont {N.~M.}\ \bibnamefont {Nor}}, \bibinfo {author} {\bibfnamefont {N.-A.}\ \bibnamefont {Rezle}}, \bibinfo {author} {\bibfnamefont {K.-W.}\ \bibnamefont {Kelvin-Lee}}, \bibinfo {author} {\bibfnamefont {M.-H.}\ \bibnamefont {Koh}}, \bibinfo {author} {\bibfnamefont {L.}~\bibnamefont {Bonneau}},\ and\ \bibinfo {author} {\bibfnamefont {P.}~\bibnamefont {Quentin}},\ }\href@noop {} {\bibfield  {journal} {\bibinfo  {journal} {Phys. Rev. C}\ }\textbf {\bibinfo {volume} {99}},\ \bibinfo {pages} {064306} (\bibinfo {year} {2019})}\BibitemShut {NoStop}%
\bibitem [{\citenamefont {Brack}\ \emph {et~al.}(1972)\citenamefont {Brack}, \citenamefont {Damgaard},\ and\ \citenamefont {\textit{et al.}}}]{FunnyHill}%
  \BibitemOpen
  \bibfield  {author} {\bibinfo {author} {\bibfnamefont {M.}~\bibnamefont {Brack}}, \bibinfo {author} {\bibfnamefont {J.}~\bibnamefont {Damgaard}},\ and\ \bibinfo {author} {\bibfnamefont {A.~J.}\ \bibnamefont {\textit{et al.}}},\ }\href@noop {} {\bibfield  {journal} {\bibinfo  {journal} {Rev. Mod. Phys.}\ }\textbf {\bibinfo {volume} {44}},\ \bibinfo {pages} {320} (\bibinfo {year} {1972})}\BibitemShut {NoStop}%
\bibitem [{\citenamefont {Brack}\ and\ \citenamefont {Pauli}(1973)}]{NPA207}%
  \BibitemOpen
  \bibfield  {author} {\bibinfo {author} {\bibfnamefont {M.}~\bibnamefont {Brack}}\ and\ \bibinfo {author} {\bibfnamefont {H.}~\bibnamefont {Pauli}},\ }\href@noop {} {\bibfield  {journal} {\bibinfo  {journal} {Nucl. Phys. A}\ }\textbf {\bibinfo {volume} {207}},\ \bibinfo {pages} {401} (\bibinfo {year} {1973})}\BibitemShut {NoStop}%
\bibitem [{\citenamefont {Moszkowski}(1957)}]{Moszkowski}%
  \BibitemOpen
  \bibfield  {author} {\bibinfo {author} {\bibfnamefont {S.~A.}\ \bibnamefont {Moszkowski}},\ }\href@noop {} {\bibfield  {journal} {\bibinfo  {journal} {Handbuch der Physik}\ }\textbf {\bibinfo {volume} {XXXIX}},\ \bibinfo {pages} {469} (\bibinfo {year} {1957})}\BibitemShut {NoStop}%
\bibitem [{\citenamefont {Ring}\ and\ \citenamefont {Schuck}(1980)}]{RingSchuck}%
  \BibitemOpen
  \bibfield  {author} {\bibinfo {author} {\bibfnamefont {P.}~\bibnamefont {Ring}}\ and\ \bibinfo {author} {\bibfnamefont {P.}~\bibnamefont {Schuck}},\ }\href@noop {} {\emph {\bibinfo {title} {The Nuclear Many-body Problem}}}\ (\bibinfo  {publisher} {Springer-Verlag},\ \bibinfo {year} {1980})\ p.\ \bibinfo {pages} {240}\BibitemShut {NoStop}%
\bibitem [{\citenamefont {Slater}(1951)}]{Slater}%
  \BibitemOpen
  \bibfield  {author} {\bibinfo {author} {\bibfnamefont {J.}~\bibnamefont {Slater}},\ }\href@noop {} {\bibfield  {journal} {\bibinfo  {journal} {Phys. Rev.}\ }\textbf {\bibinfo {volume} {81}},\ \bibinfo {pages} {385} (\bibinfo {year} {1951})}\BibitemShut {NoStop}%
\bibitem [{\citenamefont {Titin-Schnaider}\ and\ \citenamefont {Quentin}(1974)}]{Titin}%
  \BibitemOpen
  \bibfield  {author} {\bibinfo {author} {\bibfnamefont {C.}~\bibnamefont {Titin-Schnaider}}\ and\ \bibinfo {author} {\bibfnamefont {P.}~\bibnamefont {Quentin}},\ }\href@noop {} {\bibfield  {journal} {\bibinfo  {journal} {Phys. Lett. B}\ }\textbf {\bibinfo {volume} {49}},\ \bibinfo {pages} {397} (\bibinfo {year} {1974})}\BibitemShut {NoStop}%
\bibitem [{\citenamefont {Skalski}(2001)}]{Skalski}%
  \BibitemOpen
  \bibfield  {author} {\bibinfo {author} {\bibfnamefont {J.}~\bibnamefont {Skalski}},\ }\href@noop {} {\bibfield  {journal} {\bibinfo  {journal} {Phys. Rev. C}\ }\textbf {\bibinfo {volume} {63}},\ \bibinfo {pages} {024312} (\bibinfo {year} {2001})}\BibitemShut {NoStop}%
\bibitem [{\citenamefont {Bloas}\ \emph {et~al.}(2011)\citenamefont {Bloas}, \citenamefont {Koh}, \citenamefont {Quentin}, \citenamefont {Bonneau},\ and\ \citenamefont {Ithnin}}]{Bloas}%
  \BibitemOpen
  \bibfield  {author} {\bibinfo {author} {\bibfnamefont {J.~L.}\ \bibnamefont {Bloas}}, \bibinfo {author} {\bibfnamefont {M.-H.}\ \bibnamefont {Koh}}, \bibinfo {author} {\bibfnamefont {P.}~\bibnamefont {Quentin}}, \bibinfo {author} {\bibfnamefont {L.}~\bibnamefont {Bonneau}},\ and\ \bibinfo {author} {\bibfnamefont {J.}~\bibnamefont {Ithnin}},\ }\href@noop {} {\bibfield  {journal} {\bibinfo  {journal} {Phys. Rev. C}\ }\textbf {\bibinfo {volume} {84}},\ \bibinfo {pages} {0143310} (\bibinfo {year} {2011})}\BibitemShut {NoStop}%
\bibitem [{\citenamefont {Bonche}\ \emph {et~al.}(1985)\citenamefont {Bonche}, \citenamefont {Flocard}, \citenamefont {Heenen}, \citenamefont {Krieger},\ and\ \citenamefont {Weiss}}]{Bonche_pairing}%
  \BibitemOpen
  \bibfield  {author} {\bibinfo {author} {\bibfnamefont {P.}~\bibnamefont {Bonche}}, \bibinfo {author} {\bibfnamefont {H.}~\bibnamefont {Flocard}}, \bibinfo {author} {\bibfnamefont {P.}~\bibnamefont {Heenen}}, \bibinfo {author} {\bibfnamefont {S.}~\bibnamefont {Krieger}},\ and\ \bibinfo {author} {\bibfnamefont {M.}~\bibnamefont {Weiss}},\ }\href@noop {} {\bibfield  {journal} {\bibinfo  {journal} {Nucl. Phys. A}\ }\textbf {\bibinfo {volume} {443}},\ \bibinfo {pages} {39} (\bibinfo {year} {1985})}\BibitemShut {NoStop}%
\bibitem [{\citenamefont {Beiner}\ \emph {et~al.}(1975)\citenamefont {Beiner}, \citenamefont {Flocard}, \citenamefont {Giai},\ and\ \citenamefont {Quentin}}]{SIII}%
  \BibitemOpen
  \bibfield  {author} {\bibinfo {author} {\bibfnamefont {M.}~\bibnamefont {Beiner}}, \bibinfo {author} {\bibfnamefont {H.}~\bibnamefont {Flocard}}, \bibinfo {author} {\bibfnamefont {N.~V.}\ \bibnamefont {Giai}},\ and\ \bibinfo {author} {\bibfnamefont {P.}~\bibnamefont {Quentin}},\ }\href@noop {} {\bibfield  {journal} {\bibinfo  {journal} {Nucl. Phys. A}\ }\textbf {\bibinfo {volume} {238}},\ \bibinfo {pages} {29} (\bibinfo {year} {1975})}\BibitemShut {NoStop}%
\bibitem [{\citenamefont {Minkov}\ \emph {et~al.}(2022)\citenamefont {Minkov}, \citenamefont {Bonneau}, \citenamefont {Quentin}, \citenamefont {Bartel}, \citenamefont {Molique},\ and\ \citenamefont {Ivanova}}]{K_isomer}%
  \BibitemOpen
  \bibfield  {author} {\bibinfo {author} {\bibfnamefont {N.}~\bibnamefont {Minkov}}, \bibinfo {author} {\bibfnamefont {L.}~\bibnamefont {Bonneau}}, \bibinfo {author} {\bibfnamefont {P.}~\bibnamefont {Quentin}}, \bibinfo {author} {\bibfnamefont {J.}~\bibnamefont {Bartel}}, \bibinfo {author} {\bibfnamefont {H.}~\bibnamefont {Molique}},\ and\ \bibinfo {author} {\bibfnamefont {D.}~\bibnamefont {Ivanova}},\ }\href@noop {} {\bibfield  {journal} {\bibinfo  {journal} {Phys. Rev. C}\ }\textbf {\bibinfo {volume} {105}},\ \bibinfo {pages} {044329} (\bibinfo {year} {2022})}\BibitemShut {NoStop}%
\bibitem [{\citenamefont {Flocard}\ \emph {et~al.}(1973{\natexlab{a}})\citenamefont {Flocard}, \citenamefont {Quentin}, \citenamefont {Kerman},\ and\ \citenamefont {Vautherin}}]{NPA203}%
  \BibitemOpen
  \bibfield  {author} {\bibinfo {author} {\bibfnamefont {H.}~\bibnamefont {Flocard}}, \bibinfo {author} {\bibfnamefont {P.}~\bibnamefont {Quentin}}, \bibinfo {author} {\bibfnamefont {A.}~\bibnamefont {Kerman}},\ and\ \bibinfo {author} {\bibfnamefont {D.}~\bibnamefont {Vautherin}},\ }\href@noop {} {\bibfield  {journal} {\bibinfo  {journal} {Nucl. Phys. A}\ }\textbf {\bibinfo {volume} {203}},\ \bibinfo {pages} {433} (\bibinfo {year} {1973}{\natexlab{a}})}\BibitemShut {NoStop}%
\bibitem [{\citenamefont {Dobaczewski}\ \emph {et~al.}(2001)\citenamefont {Dobaczewski}, \citenamefont {Magierski}, \citenamefont {Nazarewicz}, \citenamefont {Satula},\ and\ \citenamefont {Szymanski}}]{consist12}%
  \BibitemOpen
  \bibfield  {author} {\bibinfo {author} {\bibfnamefont {J.}~\bibnamefont {Dobaczewski}}, \bibinfo {author} {\bibfnamefont {P.}~\bibnamefont {Magierski}}, \bibinfo {author} {\bibfnamefont {W.}~\bibnamefont {Nazarewicz}}, \bibinfo {author} {\bibfnamefont {W.}~\bibnamefont {Satula}},\ and\ \bibinfo {author} {\bibfnamefont {Z.}~\bibnamefont {Szymanski}},\ }\href@noop {} {\bibfield  {journal} {\bibinfo  {journal} {Phys. Rev. C}\ }\textbf {\bibinfo {volume} {63}},\ \bibinfo {pages} {024308} (\bibinfo {year} {2001})}\BibitemShut {NoStop}%
\bibitem [{\citenamefont {Duguet}\ \emph {et~al.}(2001)\citenamefont {Duguet}, \citenamefont {Bonche}, \citenamefont {Heenen},\ and\ \citenamefont {Meyer}}]{consist13}%
  \BibitemOpen
  \bibfield  {author} {\bibinfo {author} {\bibfnamefont {T.}~\bibnamefont {Duguet}}, \bibinfo {author} {\bibfnamefont {P.}~\bibnamefont {Bonche}}, \bibinfo {author} {\bibfnamefont {P.-H.}\ \bibnamefont {Heenen}},\ and\ \bibinfo {author} {\bibfnamefont {J.}~\bibnamefont {Meyer}},\ }\href@noop {} {\bibfield  {journal} {\bibinfo  {journal} {Phys. Rev. C}\ }\textbf {\bibinfo {volume} {65}},\ \bibinfo {pages} {014310} (\bibinfo {year} {2001})}\BibitemShut {NoStop}%
\bibitem [{\citenamefont {Bonneau}\ \emph {et~al.}(2015)\citenamefont {Bonneau}, \citenamefont {Minkov}, \citenamefont {Duc}, \citenamefont {Quentin},\ and\ \citenamefont {Bartel}}]{Magnet}%
  \BibitemOpen
  \bibfield  {author} {\bibinfo {author} {\bibfnamefont {L.}~\bibnamefont {Bonneau}}, \bibinfo {author} {\bibfnamefont {N.}~\bibnamefont {Minkov}}, \bibinfo {author} {\bibfnamefont {D.~D.}\ \bibnamefont {Duc}}, \bibinfo {author} {\bibfnamefont {P.}~\bibnamefont {Quentin}},\ and\ \bibinfo {author} {\bibfnamefont {J.}~\bibnamefont {Bartel}},\ }\href@noop {} {\bibfield  {journal} {\bibinfo  {journal} {Phys. Rev. C}\ }\textbf {\bibinfo {volume} {91}},\ \bibinfo {pages} {054307} (\bibinfo {year} {2015})}\BibitemShut {NoStop}%
\bibitem [{\citenamefont {\r{A} Bohr}\ and\ \citenamefont {Mottelson}(1998)}]{BM1}%
  \BibitemOpen
  \bibfield  {author} {\bibinfo {author} {\bibnamefont {\r{A} Bohr}}\ and\ \bibinfo {author} {\bibfnamefont {B.}~\bibnamefont {Mottelson}},\ }\href@noop {} {\emph {\bibinfo {title} {Nuclear Structure Vol. II}}}\ (\bibinfo  {publisher} {World Scientific, Singapore},\ \bibinfo {year} {1998})\ p.~\bibinfo {pages} {32}\BibitemShut {NoStop}%
\bibitem [{\citenamefont {Koh}\ \emph {et~al.}(2016)\citenamefont {Koh}, \citenamefont {Duc}, \citenamefont {Hao}, \citenamefont {Long}, \citenamefont {Quentin},\ and\ \citenamefont {Bonneau}}]{Koh_epja_2016}%
  \BibitemOpen
  \bibfield  {author} {\bibinfo {author} {\bibfnamefont {M.}~\bibnamefont {Koh}}, \bibinfo {author} {\bibfnamefont {D.}~\bibnamefont {Duc}}, \bibinfo {author} {\bibfnamefont {T.~N.}\ \bibnamefont {Hao}}, \bibinfo {author} {\bibfnamefont {H.}~\bibnamefont {Long}}, \bibinfo {author} {\bibfnamefont {P.}~\bibnamefont {Quentin}},\ and\ \bibinfo {author} {\bibfnamefont {L.}~\bibnamefont {Bonneau}},\ }\href@noop {} {\bibfield  {journal} {\bibinfo  {journal} {Eur. Phys. J. A}\ }\textbf {\bibinfo {volume} {52}},\ \bibinfo {pages} {3} (\bibinfo {year} {2016})}\BibitemShut {NoStop}%
\bibitem [{\citenamefont {NNDC}()}]{NNDC}%
  \BibitemOpen
  \bibfield  {author} {\bibinfo {author} {\bibnamefont {NNDC}},\ }\href@noop {} {\bibinfo  {journal} {\url{https://www.nndc.bnl.gov/nudat3/}}\ }\BibitemShut {NoStop}%
\bibitem [{\citenamefont {Raman}\ \emph {et~al.}(2001)\citenamefont {Raman}, \citenamefont {Jr.},\ and\ \citenamefont {Tikkanen}}]{BE2}%
  \BibitemOpen
\bibfield  {journal} {  }\bibfield  {author} {\bibinfo {author} {\bibfnamefont {S.}~\bibnamefont {Raman}}, \bibinfo {author} {\bibfnamefont {C.~N.}\ \bibnamefont {Jr.}},\ and\ \bibinfo {author} {\bibfnamefont {P.}~\bibnamefont {Tikkanen}},\ }\href@noop {} {\bibfield  {journal} {\bibinfo  {journal} {At. Data Nucl. Tables}\ }\textbf {\bibinfo {volume} {78}},\ \bibinfo {pages} {1} (\bibinfo {year} {2001})}\BibitemShut {NoStop}%
\bibitem [{\citenamefont {NUMOR}()}]{Spect}%
  \BibitemOpen
  \bibfield  {author} {\bibinfo {author} {\bibnamefont {NUMOR}},\ }\href@noop {} {\bibinfo  {journal} {\url{https://magneticmoments.info/numor/in_search.php} (cut-off date 2019.03.31) and references quoted therein}\ }\BibitemShut {NoStop}%
\bibitem [{\citenamefont {Flocard}\ \emph {et~al.}(1973{\natexlab{b}})\citenamefont {Flocard}, \citenamefont {Quentin},\ and\ \citenamefont {Vautherin}}]{PLB}%
  \BibitemOpen
\bibfield  {journal} {  }\bibfield  {author} {\bibinfo {author} {\bibfnamefont {H.}~\bibnamefont {Flocard}}, \bibinfo {author} {\bibfnamefont {P.}~\bibnamefont {Quentin}},\ and\ \bibinfo {author} {\bibfnamefont {D.}~\bibnamefont {Vautherin}},\ }\href@noop {} {\bibfield  {journal} {\bibinfo  {journal} {Phys. Lett. B}\ }\textbf {\bibinfo {volume} {46}},\ \bibinfo {pages} {304} (\bibinfo {year} {1973}{\natexlab{b}})}\BibitemShut {NoStop}%
\end{thebibliography}%

\end{document}